\documentclass[aps,twocolumn,pra,superscriptaddress,showpacs,tightenlines]{revtex4}
\usepackage{amssymb}
\usepackage{amsmath}
\usepackage{graphicx}
\usepackage{epsfig}
\usepackage{txfonts}
\usepackage{subfigure}
\usepackage{amsfonts}
\usepackage{CJK}

\begin{document}

\begin{CJK*}{GBK}{song}
\title{Quantum discord amplification induced by quantum phase transition via a cavity-Bose-Einstein-condensate system}
\author{Ji-Bing Yuan and Le-Man Kuang\footnote{Author to whom any correspondence should be
addressed. }\footnote{ Email: lmkuang@hunnu.edu.cn}}

\affiliation{Key Laboratory of Low-Dimensional Quantum Structures
and Quantum Control of Ministry of Education,  and Department of
Physics, Hunan Normal University, Changsha 410081, China}
\date{\today}

\begin{abstract}
We propose a theoretical scheme to realize a sensitive amplification
of quantum discord (QD) between two atomic qubits via a
cavity-Bose-Einstein condensate (BEC) system which was used to
firstly realize the Dicke quantum phase transition (QPT) [Nature
\textbf{464}, 1301 (2010)]. It is shown that the influence of the
cavity-BEC system upon the two qubits is equivalent to a phase
decoherence environment. It is found that QPT in the cavity-BEC
system is the physical mechanism of the sensitive QD amplification.
\end{abstract}

\maketitle \narrowtext
\end{CJK*}

Quantum discord (QD) ~\cite{Ollivier2001,Henderson2003} is
considered to be a more general resource than quantum entanglement
in quantum information
processing~\cite{Datta2008,Lanyon2008,Roa2011,Li2012,Datta2009,Madhok2012}.
A nonzero QD in some separable states is responsible for the quantum
computational efficiency of deterministic quantum computation with
one pure qubit~\cite{Datta2008,Lanyon2008,Knill1998} and also has
been considered as an useful resource in quantum locking
~\cite{Datta2009} and quantum state
discrimination~\cite{Roa2011,Li2012}. On the other hand, any
realistic quantum systems interact inevitably with their surrounding
environments, which introduce quantum noise into the systems. As an
useful resource, it is interesting that QD can be amplified by the
quantum noise. We find the QD does be amplified for two
non-interacting qubits immersed in a common phase decoherence
environment~\cite{Yuan2010}. Especially, when the two qubits are
identical, the phase decoherence can induce a stable amplification
of the initially-prepared QD for certain $X$-type states. In this
paper, we propose a scheme to realize the controllable QD
amplification of two atomic qubits by making use of an artificial
phase decoherence environment consisting of a cavity-Bose-Einstein
condensate (BEC) system.

The Dicke model \cite{Dicke1954,EmaryC2003} describes a larger
number of two-level atoms interacting with a single cavity field
mode. As increasing atom-filed coupling, the model predicts a QPT
\cite{Sachdev1999} from the normal phase, which the atoms are in the
ground state associated with vacuum field state, to the
super-radiant phase, which both the atoms and field have collective
excitations. Recently, a  cavity-Bose-Einstein condensate (BEC)
system is employed to firstly realize the Dicke quantum phase
transition (QPT) experimentally and to explore symmetry breaking at
the Dicke QPT ~\cite{Baumann2010}. Meanwhile, the QPT system usually
displays ultra-sensitivity in its dynamical evolution near the
quantum critical point
\cite{Hepp1973,Wang1973,Emary2003,Quan2006,Huang2009}, which has
been confirmed by a NMR experiment \cite{Zhang2008}. The purpose of
this paper is to show that the QD of two initially correlated atomic
qubits can be sensitively amplified via the cavity-BEC system near
the critical point. We show that the cavity-BEC system can form an
artificial phase decoherence environment for the two atomic qubits,
and the QD of the two atomic qubits can be amplified by adjusting
the QPT parameter of the cavity-BEC system.

\begin{figure}[tbp]
\includegraphics[bb=20 410 400 785, height=5.2cm, width=2.8 in]{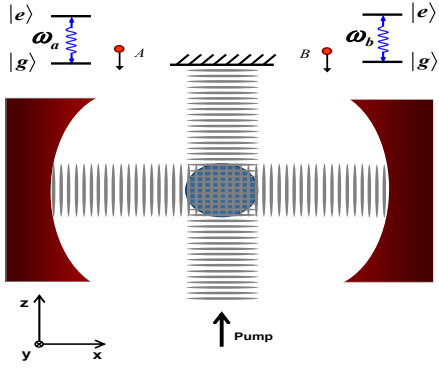}
\caption{(Color online) Schematic of our physical system: Two
 atomic qubits $A$ and $B$ with energy separation $\omega_{A}$ and
$\omega_{B}$ are, respectively, injected into the cavity in which a
  atomic BEC couples to a single cavity field and a transverse pump field.}
\label{fig1.eps}
\end{figure}

The physical system under our consideration is shown in
Fig.~\ref{fig1.eps}. A BEC with $N$ identical two-level $^{87}Rb$
atoms is confined in a ultrahigh-finesse optical cavity. The atoms
interact with a single cavity model of frequency $\omega_{c}$ and a
transverse pump field of frequency $\omega_{p}$. We consider a
situation that the frequency $\omega_{c}$ and $\omega_{p}$ are
detuned far from the atomic resonance frequency $\omega_{A}$ so that
the dutunings far exceed the rate of atomic spontaneous emission,
the atoms only scatter photons either along or transverse to the
cavity axis. Before the pump field turns on, atoms in the BEC are
supposed to be in the zero-momentum state
$|p_{x},p_{z}\rangle=|0,0\rangle$. As soon as one turns on the pump
field, via the photon scattering of the pump and cavity fields, some
atoms are excited into the momentum sates $|p_{x},p_{z}\rangle=|\pm
k,\pm k\rangle=\sum_{\upsilon_{1},\upsilon_{2}=\pm1}|\upsilon_{1} k,
\upsilon_{2}  k\rangle$ (hereafter, we take $\hbar=1$.) due to the
conservation of momentum, where $k$ is the wave-vector, which is
approximated to be equal on the cavity and pump fields.  The two
momentum states with $|0,0\rangle$ and $|\pm k,\pm k\rangle$ are
regarded as two levels of the $^{87}Rb$ atom with energy separation
$\omega_{0}=k^{2}/m$, where $m$ is the mass of $^{87}Rb$ atom.
Defined the collective operators $\hat{J_{z}}=\sum_{i}|\pm k,\pm
k\rangle \langle \pm k,\pm k|$,
$\hat{J_{+}}=\hat{J_{-}^{\dag}}=\sum_{i} |\pm k,\pm k \rangle
\langle 0,0 |$  with the index $i$ labelling the atoms, the cavity-
BEC is described by the Dick model ~\cite{Baumann2010, Baumann2011}
\begin{equation}
\label{1}\hat{H}_{1}=\omega\hat{a}^{\dag}\hat{a}+\omega_{0}\hat{J}_{z}+\frac{\lambda
}{\sqrt{N}}\left(\hat{a}+\hat{a}^{\dag}\right)(\hat{J_{+}}+\hat{J_{-}}),
\end{equation}
where $\hat{a}^{\dag}(\hat{a})$ is the creation (annihilation)
operator of the cavity field. $\omega=-\Delta_{c}+ U_{0}N/2$ is
effective frequency of the cavity field including the frequency
shift induced by the BEC under the  frequency of pump field rotating
frame, where $\Delta_{c}=\omega_{p}- \omega_{c}$ is the dutuning
between pump field and cavity field and $U_{0}=g_{0}^{2}/\Delta$ is
the frequency shift of a single atom with maximally cavity field
coupling strength $g_{0}$ and detuning
$\Delta=\omega_{p}-\omega_{A}$.
$\lambda=\sqrt{N}g_{0}\Omega_{p}/2\Delta$ is the coupling strength
induced by the cavity field and pump field with $\Omega_{p}$
denoting the maximum pump Rabi frequency which can be adjusted by
the pump power.

We consider such a situation that the two atomic qubits pass through
the cavity at the same time and interact with the single cavity
field, the expression of the Hamiltonian reads as
\begin{equation}
\hat{H}_{2}=\omega\hat{a}^{\dag}\hat{a}+\frac{\omega_{A}}{2}\hat{\sigma}_{z}^{A}
+\frac{\omega_{B}}{2}\hat{\sigma}_{z}^{B}+\left(g_{A}\hat{a}^{\dag}\hat{\sigma}_{-}^{A}
+g_{B}\hat{a}^{\dag}\hat{\sigma}_{-}^{B}+H.c\right), \label{2}
\end{equation}
where  $\hat{\sigma}^{A (B)} _{z}=\left\vert e\right\rangle
\left\langle e\right\vert -\left\vert g\right\rangle \left\langle
g\right\vert$ is pauli operator with $ \left\vert e\right\rangle$
and $\left\vert g\right\rangle$ being the excited and ground states.
$\sigma^{A (B)}_{+} (\sigma^{A (B)}_{-})$ is the raising
operator(lowering operator). $g_{A (B)}$ is the coupling strength
between the atomic qubit A(B) and the cavity field, $\omega_{A (B)}$
is the energy separation. Here we have made a rotating wave
approximation. If the atomic qubit is far-off-resonant with the
cavity field satisfying the detuning $\Delta_{A (B)}=\omega_{A
(B)}-\omega$ is much large than the corresponding coupling coupling
strength $g_{A(B)}$, one can use the
Fr$\ddot{\mathrm{o}}$hlich-Nakajima transformation
\cite{Fr1950,Naka1955} to make the Hamiltonian in Eq. (\ref{2})
become the following expression
\begin{equation}
\hat{H}_{2}^{^{\prime
}}=\frac{1}{2}\omega_{A}^{^{\prime}}\hat{\sigma}_{z}^{A}
+\frac{1}{2}\omega_{B}^{^{\prime}}\hat{\sigma}_{z}^{B}+\left(\omega
+\delta_{A}\hat{\sigma}_{z}^{A}+\delta_{B}\hat{\sigma}_{z}^{B}\right)
\hat{a}^{\dag }\hat{a},
\end{equation}
where $\omega_{A(B)}^{^{\prime}}=\omega_{A(B)}+\delta_{A(B)}$ with
$\delta_{A(B)} =g^{2}_{A(B)}/\Delta_{A(B)}$ being the frequency
shift induced by the scattering between cavity field and atomic
qubit A(B). Then the effective Hamiltonian describing the two atomic
qubits passing through the cavity-BEC system is
\begin{equation}
\label{4}
\hat{H}_{eff}=\frac{1}{2}\omega_{A}^{^{\prime}}\hat{\sigma}_{z}^{A}+\frac{1
}{2}\omega_{B}^{^{\prime}}\hat{\sigma}_{z}^{B}+\left(\delta_{A}\hat{\sigma}_{z}^{A}
+\delta_{B}\hat{\sigma}_{z}^{B}\right)\hat{a}^{\dag}\hat{a}+\hat{H}_{1}.
\end{equation}

Now we consider the dynamics of the two atomic qubits passing
through the cavity-BEC system. We assume the two atomic qubits are
initially prepared in a class of state with maximally mixed
marginals ($\hat{\rho} ^{A(B)}=\hat{I}^{A(B)}/2$) described by the
three-parameter $X$-type density matrix
$\hat{\rho}_{s}(0)=1/4(\hat{I}^{AB}+\sum_{i=1}^{3}c_{i}\hat{\sigma}_{i}^{A}\otimes
\hat{\sigma} _{i}^{B})$, where $\hat{I}^{AB}$ is the identity
operator in the Hilbert space of the two atomic qubits, $i=1,2,3$
mean $x,y,z$ correspondingly, and $c_{i}$ ($0\leq \left\vert
c_{i}\right\vert \leq 1$) are real numbers satisfying the unit trace
and positivity conditions of the density operator
$\hat{\rho}_{s}(0)$. The cavity-BEC system is initially in the
ground state $|G\rangle$ of the Hamiltonian in Eq. (\ref{1}). The
dynamic evolution of the total system is controlled by the
Hamiltonian in Eq. (\ref{4}). The density operator of the total
system at time $t$ is written as
$\rho_{T}(t)=U(\hat{\rho}_{s}(0)\otimes|G\rangle\langle G|)U^{\dag}$
with $U=e^{-i\hat{H}_{eff}t}$. After tracing the degree of freedom
of the cavity-BEC system, we obtain the reduced density of the two
atomic qubits
\begin{equation}
\label{6} \hat{\rho}_{s}(t)=\frac{1}{4}\left(
\begin{array}{cccc}
1+c_{3} & 0 & 0 & \mu(t)D_{1}(t) \\
0 & 1-c_{3} & \nu(t)D_{2}(t) & 0 \\
0 & \nu^{*}(t)D^{*}_{2}(t) & 1-c_{3} & 0 \\
\mu^{*}(t)D^{*}_{1}(t) & 0 & 0 & 1+c_{3}
\end{array}
\right),
\end{equation}
where we have introduced the following parameters
\begin{eqnarray}
\mu(t)&=&(c_{1}-c_{2})e^{-i(\omega^{'}_{A}+\omega^{'}_{B})t},
\hspace{0.2cm} \nu(t) =(c_{1}+c_{2})e^{-i(\omega^{'}_{A}-\omega^{'}_{B})t},\nonumber \\
D_{1}(t)&=&\langle
G|e^{i\hat{H}_{ee}t}e^{-i\hat{H}_{gg}t}|G\rangle,\hspace{0.2cm}D_{2}(t)=\langle
G|e^{i\hat{H}_{eg}t}e^{-i\hat{H}_{ge}t}|G\rangle,\label{7}
\end{eqnarray}
with
\begin{eqnarray}
\hat{H}_{ee}&=&\delta_{1}\hat{a}^{\dag}\hat{a}+\hat{H}_{1},
\hat{H}_{gg}=-\delta_{1}\hat{a}^{\dag}\hat{a}+\hat{H}_{1},\delta_{1}=\delta_{A}+\delta_{B}\nonumber\\
\hat{H}_{eg}&=&\delta_{2}\hat{a}^{\dag}\hat{a}+\hat{H}_{1},
\hat{H}_{ge}=-\delta_{2}\hat{a}^{\dag}\hat{a}+\hat{H}_{1},\delta_{2}=\delta_{A}-\delta_{B}.\label{8}
\end{eqnarray}

We consider the situation that two atomic qubits pass through the
cavity field region in a very short time of satisfying the
conditions $\delta_{1}t\ll1$ and $\delta_{2}t\ll1$. In fact,
according to Ref. \cite{Baumann2010}, the waist of the cavity field
is 25 $\mu m$, the effective frequency shift $\delta_{A}$,
$\delta_{B}$ are about 100 Hz, above conditions are well satisfied
if injected velocity of the atomic qubits meets $v\gg10^{-3}$  m/s.
By the short time approximation, the factors $|D_{1}(t)|$,
$|D_{2}(t)|$ can be derived as
\begin{equation}
|D_{1}\left(t\right)|=\exp\left(-2\gamma\delta_{1}^{2}t^{2}\right)
,|D_{2}\left(t\right)|=\exp\left(-2\gamma
\delta_{2}^{2}t^{2}\right),  \label{9}
\end{equation}
where  the decay factor $\gamma
=\left\langle\left(\hat{a}^{\dag}\hat{a}\right)
^{2}\right\rangle-\left\langle\hat{a}^{\dag}\hat{a}\right\rangle^{2}$
is the cavity photon number fluctuation (PNF) in the ground state
$|G\rangle$ \cite{Huang2009}.

From Eq. (5) we can see that the cavity-BEC system only affects
off-diagonal elements of the density for the two atomic qubits,
hence it is equivalently a phase decoherence environment for the two
atomic qubits. That is, the cavity-BEC system constitutes an
artificial phase decoherence environment of the two qubits. The QPT
parameter of the cavity-BEC system $\lambda$ is a controllable
parameter of the artificial environment. It's worth noting that when
the effective frequency shift $\delta_{A}$, $\delta_{B}$ are equal,
i.e, $\delta_{2}=0$, a decoherence free space in the basis
$\{|ge\rangle, |eg\rangle\}$ appears.

In order to obtain the detailed form of the PNF $\gamma$, in the
following we give the ground state $|G\rangle$ according to the Ref.
\cite{EmaryC2003}. Utilizing the Holstein-Primakoff transformation
\cite{Holstein1940}  $\hat{J}_{+} =\hat{c}^{\dag
}\sqrt{2j-\hat{c}^{\dag }\hat{c}}, \hat{J}_{-}
=\sqrt{2j-\hat{c}^{\dag }\hat{c}} \hat{c}, \hat{J}_{z}
=\hat{c}^{\dag}\hat{c}-j$, where$j=N/2$ , the Hamiltonian of the Eq.
(\ref{1}) is further reduce to
\begin{equation}
\label{11}
\hat{H}_{1}=\omega\hat{a}^{\dag}\hat{a}+\omega_{0}\hat{c}^{\dag
}\hat{c}+\lambda\left( \hat{a}+\hat{a}^{\dag }\right)
\left(\hat{c}^{\dag}\sqrt{1-\frac{\hat{c}^{\dag}\hat{c}}{2j}}+H.c.\right).
\end{equation}
When the coupling strength $\lambda$ is smaller than the critical
coupling strength $\lambda_{c}=\sqrt{\omega\omega_{0}}/2$, i.e.,
$\lambda<\lambda_{c}$, the system is in the normal phase where the
BEC and the cavity field have low excitations. While when the
coupling strength is larger than the critical strength, i.e.,
$\lambda>\lambda_{c}$, the system is in the super-radiant phase
where both the BEC and the cavity field have collective excitations
in the order of the atom number $N$.

In the normal phase at the thermodynamic limit $j\rightarrow\infty$,
neglecting terms with $j$ in the denominator, Hamiltonian (\ref{11})
becomes
\begin{equation}
\label{12}\hat{H}_{1}=\omega\hat{a}^{\dag}\hat{a}+\omega_{0}\hat{c}^{\dag}\hat{c}
+\lambda\left(\hat{a}+\hat{a}^{\dag}\right)\left(\hat{c}+\hat{c}^{\dag}\right),
\end{equation}
where we omit the constant term.  The Hamiltonian in Eq. (\ref{12})
can be diagonalized as
\begin{equation}
\label{13}\hat{H}_{1}=\varepsilon _{-}\hat{d}_{1}^{\dag
}\hat{d}_{1}+\varepsilon _{+}\hat{d}_{2}^{\dag }\hat{d}_{2}
\end{equation}
by the Bogoliubov transformation
\begin{eqnarray}
\label{14}\hat{a}^{\dag
}&=&f_{1}\hat{d}_{1}^{\dag}+f_{2}\hat{d}_{1}+f_{3}\hat{d}_{2}^{\dag}
+f_{4}\hat{d}_{2},\\\hat{c}^{\dag}&=&h_{1}\hat{d}_{1}^{\dag}
+h_{2}\hat{d}_{1}+h_{3}\hat{d}_{2}^{\dag}+h_{4}\hat{d}_{2}\nonumber,
\end{eqnarray}
where the eigenfrequencies $\varepsilon _{-}$ and $\varepsilon _{+}$
of the cavity-BEC system have the following expression
\begin{eqnarray}
\label{15}\varepsilon_{\pm}^{2}&=&\frac{1}{2}\left[\omega^{2}
+\omega_{0}^{2}\pm\sqrt{\left( \omega_{0}^{2}-\omega ^{2}\right)
^{2}+16\lambda ^{2}\omega \omega_{0}}\right] .
\end{eqnarray}
The coefficients of Bogoliubov transformation about the cavity field
in the normal phase are
\begin{equation}
\label{16} f_{1,2}=\frac{1}{2}\frac{\cos\phi}{\sqrt{\varepsilon
_{-}\omega}}\left(\omega \pm \varepsilon _{-}\right), \hspace{0.2cm}
f_{3,4} =\frac{1}{2}\frac{\sin \phi }{\sqrt{\varepsilon _{+}\omega
}}\left(\omega \pm \varepsilon _{+}\right),
\end{equation}
where the mixing angle $\phi$ is given by
$\tan2\phi=\frac{4\lambda\sqrt{\omega\omega_{0}}}{\omega_{0}^{2}-\omega^{2}}$.

In the supper-radiant phase, we displace the bosonic modes
$\hat{a}^{\dag}\rightarrow \hat{a}^{\dag^{\prime}}+\sqrt{\alpha },
 \hat{c} ^{\dag }\rightarrow \hat{c}^{\dag^{\prime }}-\sqrt{\beta}$
with $\sqrt{\alpha}$ and $\sqrt{\beta}$ describing the macroscopic
mean fields above $\lambda_{c}$ in the order of $O(j)$. Neglecting
terms with $j$ in the denominator and taking
$\sqrt{\alpha}=\frac{2\lambda}{\omega}\sqrt{\frac{j}{2}\left(1-\xi
^{2}\right)}$, $\sqrt{\beta}=\sqrt{j\left(1-\xi \right)}$ with $\xi
=\frac{\lambda_{c}^{2}}{\lambda^{2}}$ , the Hamiltonian Eq.
(\ref{11}) is reduced to the following form
\begin{eqnarray}
\label{19} \hat{H}_{1}&=&\omega\hat{a}^{\dag^{\prime
}}\hat{a}^{^{\prime}}+\tilde{\omega }_{0}\hat{c}^{\dag
^{\prime}}\hat{c}^{^{\prime}}+\eta \left(\hat{c}^{\dag ^{\prime
}}+\hat{c}^{^{\prime}}\right)^{2}\nonumber\\
&&+\tilde{\lambda}\left(\hat{a}^{^{\prime}}+\hat{a}^{\dag^{\prime
}}\right)\left(\hat{c}^{\dag^{\prime}}+\hat{c}^{^{\prime}}\right),
\end{eqnarray}
where the parameters $\tilde{\omega}_{0}$, $\tilde{\lambda}$ and
$\eta$ are given by
\begin{eqnarray}
\label{20}\tilde{\omega}_{0}&=&\frac{\omega_{0}}{2\xi}(1+\xi)
,\hspace{0.5cm}\tilde{\lambda}=\lambda \xi \sqrt{\frac{2}{1+\xi}}, \nonumber\\
\eta&=&\frac{\omega_{0}\left(1-\xi\right)\left(3+\xi\right)}{8\xi\left(1+\xi\right)}.
\end{eqnarray}

The Hamiltonian in Eq. (\ref{19}) also can be diagonalized as
\begin{eqnarray}
\label{21}\hat{H}&=&\varepsilon _{-}^{^{\prime
}}\hat{d}_{1}^{^{\prime }\dag }\hat{d} _{1}^{^{\prime }}+\varepsilon
_{+}^{^{\prime }}\hat{d}_{2}^{^{\prime }\dag}\hat{d}_{2}^{^{\prime}}
\end{eqnarray}
by the Bogoliubov transformation
\begin{eqnarray}
\label{22}\hat{a}^{\dag^{\prime}}&=&f_{1}^{^{\prime}}\hat{d}_{1}^{\dag
^{\prime }}+f_{2}^{^{\prime}}\hat{d}_{1}^{^{\prime}}+f_{3}^{^{\prime
}}\hat{d}_{2}^{\dag ^{\prime }}+f_{4}^{^{\prime}}\hat{d}_{2}^{^{\prime}},\\
\hat{c}^{\dag ^{\prime }}&=&h_{1}^{^{\prime}}\hat{d}_{1}^{\dag
^{\prime}}+h_{2}^{^{\prime}}\hat{d}_{1}^{^{\prime}}+h_{3}^{^{\prime
}}\hat{d}_{2}^{\dag^{\prime}}+h_{4}^{^{\prime
}}\hat{d}_{2}^{^{\prime}}\nonumber,
\end{eqnarray}
where the eigenfrequencies $\varepsilon_{-}^{'}$ and
$\varepsilon_{+}^{'}$ read as
\begin{eqnarray}
\label{23}\varepsilon_{\pm}^{^{\prime }2}=\frac{1}{2}\left[\omega
^{2}+\frac{\omega_{0}^{2}}{\xi^{2}} \pm\sqrt{\left(\omega
^{2}-\frac{\omega_{0}^{2}}{\xi^{2}}\right)^{2} +4\omega^{2}
\omega}_{0}^{2}\right].
\end{eqnarray}
The coefficients of Bogoliubov transformation about the cavity field
in the super-radiant phase are
\begin{equation}
\label{24}f_{1,2}^{^{\prime }}=\frac{1}{2}\frac{\cos \phi ^{^{\prime
}}}{\sqrt{ \varepsilon _{-}^{^{\prime }}\omega }}\left( \omega \pm
\varepsilon_{-}^{^{\prime }}\right), \hspace{0.3cm}
f_{3,4}^{^{\prime}}=\frac{1}{2}\frac{\sin\phi ^{^{\prime }}}{\sqrt{
\varepsilon _{+}^{^{\prime }}\omega }}\left( \omega \pm \varepsilon
_{+}^{^{\prime }}\right) .
\end{equation}
where $\phi^{'}$ is the mixing angle defined by $\tan 2\phi
^{^{\prime }}=\frac{2\omega
\omega_{0}\xi^{2}}{\omega_{0}^{2}-\xi^{2}\omega^{2}}$.

The PNF can be given in the normal phase with ground state
$|0,0\rangle_{d_{1},d_{2}}$ and  the super-radiant phase  with
ground state $|0,0\rangle_{d_{1}^{'},d_{2}^{'}}$, respectively, as
the following forms
\begin{eqnarray}
\label{26} \gamma &=&\left\{
\begin{array}{c}
\vspace{0.5cm} 2f_{1}^{2}f_{2}^{2}+2f_{3}^{2}f_{4}^{2}+\left(
f_{1}f_{4}+f_{2}f_{3}\right) ^{2},
\hspace{0.3cm}\lambda<\lambda_{c}, \\
\vspace{0.1cm}f_{1}^{^{\prime }2}f_{2}^{^{\prime
}2}+2f_{3}^{^{\prime }2}f_{4}^{^{\prime }2}+\left( f_{1}^{^{\prime
}}f_{4}^{^{\prime}}+f_{2}^{^{\prime }}f_{3}^{^{\prime}}\right) ^{2}
\hspace{1.2cm}\\+\alpha\left[\left(
f_{1}^{^{\prime}}+f_{2}^{^{\prime }}\right) ^{2}+\left(
f_{3}^{^{\prime }}+f_{4}^{^{\prime
}}\right)^{2}\right],\hspace{0.8cm}\lambda>\lambda_{c}.
\end{array}
\right.
\end{eqnarray}

\begin{figure}[tbp]
\includegraphics[height=4.2cm, width=2.8in]{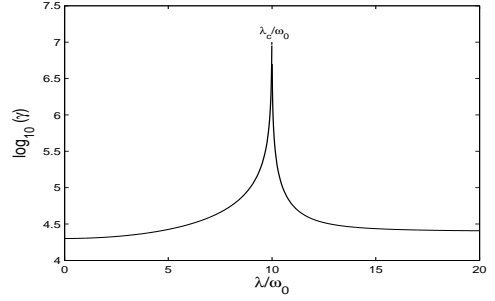}
\caption{(Color online)  The logarithm to base 10 of the PNF
$\gamma$\ changes with  the coupling strength $\lambda$. Related
parameters $N=10^5$, $\omega_{0}=0.05$MHz,
 $\omega=20$MHz correspond the experimental parameters in
Ref. \cite{Baumann2010}.} \label{fig2.eps}
\end{figure}

Compared with the case of the normal phase, the displacement
$\alpha$ due to collective excitation appears in the super-radiant
phase. Figure \ref{fig2.eps} shows the PNF $\gamma$ will experience
drastic change near the critical coupling point
$\lambda_{c}/\omega_{0}=10$. The closer the coupling strength
$\lambda$  near the critical coupling point, the larger the PNF
$\gamma$. This inspires us to control the coherence decay rate of
the two atomic qubits by adjusting the pumping power to change the
coupling strength in the region near the critical coupling.

In the following we consider the QD amplification of  the two atomic
qubits induced by the QPT of the cavity-BEC system. The QD
~\cite{Ollivier2001} is defined as the difference between the total
correlation and the classical correlation with the expression
$\label{28} \mathcal {D}\left(\hat{\rho}^{AB}\right)
=\mathcal{I}\left(\hat{\rho}^{A}:\hat{\rho}^{B}\right)
-\mathcal{C}\left(\hat{\rho}^{AB}\right)$ with $\hat{\rho}^{A}$,
$\hat{\rho}^{B}$, and $\hat{\rho}^{B}$ being  the reduced density
operators for subsystems $A$ and $B$, and the total density
operator, respectively. The total correlation in the state
$\hat{\rho}^{AB}$ is measured by quantum mutual information
$\mathcal{I}\left(\hat{\rho}^{A}:\hat{\rho}^{B}\right)
=S\left(\hat{\rho}^{A}\right) +S\left( \hat{\rho}^{B}\right)
-S\left(\hat{\rho}^{AB}\right)$ with $S\left(\hat{\rho}\right)
=-\textrm{Tr}(\hat{\rho} \log \hat{\rho})$ being the von Neumann
entropy. The classical correlation between the two subsystems $A$
and $B$ is given by $ C(\hat{\rho}^{AB})=S(\hat{\rho}
_{A})-\min_{\{\hat{P}^{B}_{k}\}}\left[
\sum_{k}p_{k}S(\hat{\rho}^{A}_{k})\right]$ where
$p_{k}=\textrm{Tr}_{AB}[(\hat{I}^{A}\otimes
\hat{P}^{B}_{k})\hat{\rho}^{AB}(\hat{I}^{A}\otimes
\hat{P}^{B}_{k})]$ denotes the probability relating to the outcome
$k$, and $\hat{I}^{A}$ denotes the identity operator for the
subsystem $A$ with $\{\hat{P}^{B}_{k}\}$ being a set of projects
performed locally on the subsystem $B$.

The mutual information of the state given in Eq.~(\ref{6}) is
derived as $\mathcal{I}\left(\hat{\rho}^{A}:\hat{\rho}^{B}\right)\
=2+{\sum_{i=1}^{4}}\lambda_{i}\log\lambda _{i}$, where
$\lambda_{1,2}=\frac{1}{4}(1+c_{3}\pm |\mu(t)D_{1}(t)|)$,
$\lambda_{3,4}=\frac{1}{4}(1-c_{3}\pm|\nu(t)D_{2}(t)|)$ are four
eigenvalues of $\hat{\rho}_{s}(t)$. And the classical correlation
can be obtained as \cite{Luo2008,Yuan2010} $C(\hat{\rho}_{s}(t))
=\overset{2}{\underset{n=1}{\sum }}\frac{1+(-1)^{n}\chi }{ 2}\log
_{2}\left[ 1+(-1)^{n}\chi \right]$ with
$\chi(t)=\max\left[|c_{3}|,(|\mu(t)D_{1}(t)|+|\nu(t)D_{2}(t)|)/2\right]$.
Therefore, the QD can be written as
\begin{eqnarray}
\label{34} \mathcal{D}\left(\hat{\rho}_{s}(t)\right) \
&=&2+{\sum_{i=1}^{4}}\lambda _{i}\log _{2}\lambda
_{i}-C(\hat{\rho}_{s}(t)).
\end{eqnarray}

\begin{figure}[tbp]
\includegraphics[height=5.2cm, width=2.8 in]{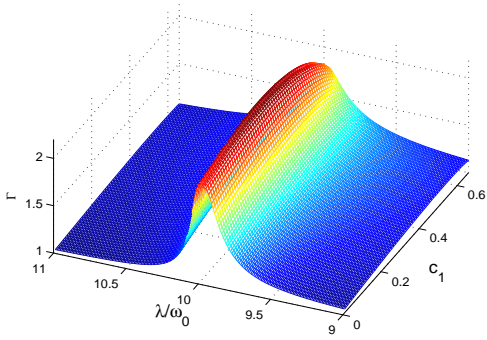}
\caption{(Color online)  The QD amplification rate as a function of
the coupling strength $\lambda$ and the initial parameter $c_{1}$.
Other parameters are set as $c_{3}=c_{1}/2$, $c_{2}=0$,
$\omega_{0}=0.05$ MHz, $\omega=20$ MHz, $t_{f}=1/\omega_{0}$,
$\delta_{1}=0.001\omega_{0}$, $\delta_{2}=0$, and $N=10^5$.}
\label{fig3.eps}
\end{figure}

The QD can be amplified for some initial states such as the state
parameters being set as $ c_{2}=0$, $0\leq c_{1}=2c_{3} \leq 2/3$
when the qubits are in the phase decoherence environment
\cite{Yuan2010}. For the present cavity-BEC environment, let the two
atomic qubits enter  the cavity at time $t=0$ and leave the cavity
at time $t_{f}$. Then we can define the QD amplification rate as
$\Gamma=\mathcal{D}(t_{f})/\mathcal{D}(0)$. In Figure \ref{fig3.eps}
we have plotted the amplification rate $\Gamma$ with respect to the
coupling strength $\lambda$ and the initial state parameter $c_{1}$
when $c_{3}=c_{1}/2$, $c_{2}=0$, $\omega_{0}=0.05$ MHz, $\omega=20$
MHz, $\lambda_c/\omega_0=10$, $t_{f}=1/\omega_{0}$,
$\delta_{1}=0.001\omega_{0}$, $\delta_{2}=0$, and $N=10^5$. Figure
\ref{fig3.eps} indicates that the initial QD can be amplified by the
use of the cavity-BEC system through changing the QPT parameter
$\lambda$. Specially, the QD amplification rate sensitively
increases at the QPT point of the cavity-BEC system
$\lambda=\lambda_c$. In this sense, the sensitive QD amplification
can be understood as a quantum phenomenon induced by the QPT of the
cavity-BEC system. It should be pointed out that one can control the
QPT parameter $\lambda$ by changing the Rabi frequency of the pump
field $\Omega_{p} $due to the relation
$\lambda=\sqrt{N}g_{0}\Omega_{p}/2\Delta$.

In conclusion, we have proposed a scheme to realize the sensitive QD
amplification of two atomic qubits via the cavity-BEC system through
by changing the QPT parameter of the the cavity-BEC system, and and
revealed the QPT mechanism of  the sensitive QD amplification. We
have indicated that the cavity-BEC system is equivalent to a phase
decoherence environment for the two atomic qubits. Hence, it
provides an artificial and controllable phase decoherence
environment for quantum information processing. It should mentioned
that the present scheme should be within the reach of present-day
techniques since the cavity-BEC system used in the scheme has been
well established in recent experiments of observing the Dicke QPT
\cite{Baumann2010}. The experimental realization of the scheme
proposed in the present paper deserves further investigation.

\acknowledgments This work was supported by the NFRP under Grant No.
2007CB925204, the NSF under Grant No. 11075050,  and the PCSIRTU
under Grant No. IRT0964, and the HPNSF under Grant No. 11JJ7001.

\end{document}